\documentclass[journal=jacsat,manuscript=article,hyperref]{achemso}

\usepackage[version=3]{mhchem} 

\usepackage{hyperref}
\usepackage{titlesec}

\hypersetup{
	colorlinks=true,
	citecolor=blue,
	linkcolor=blue,
	urlcolor=blue
}

\usepackage[none]{hyphenat} 
\SectionNumbersOn

\setkeys{acs}{doi=true}

\newcommand{\sub}[1]{_{\mathrm{#1}}} 

\author{Livin Paul}
\author{Alona Vidishcheva}
\author{Gal Sandik}
\author{Adina Golombek}
\phone{+123 (0)123 4445556}
\fax{+123 (0)123 4445557}
\affiliation[Unknown University]
{Physical Chemistry Department, Raymond \& Beverly Sackler Faculty of Exact Sciences and Center for Light–Matter Interaction, Tel Aviv University, Tel Aviv, Israel.}
\author{Tal Schwartz}
\email{talschwartz@tau.ac.il}

\title
{Cavity-Mediated Long-Range Cooperative Coupling between Localized Plasmons and Molecular Excitons}

\abbreviations{IR,NMR,UV}

\begin{document}
	
	\begin{center}
		\vspace*{1em}
		\includegraphics[height=1.75in]{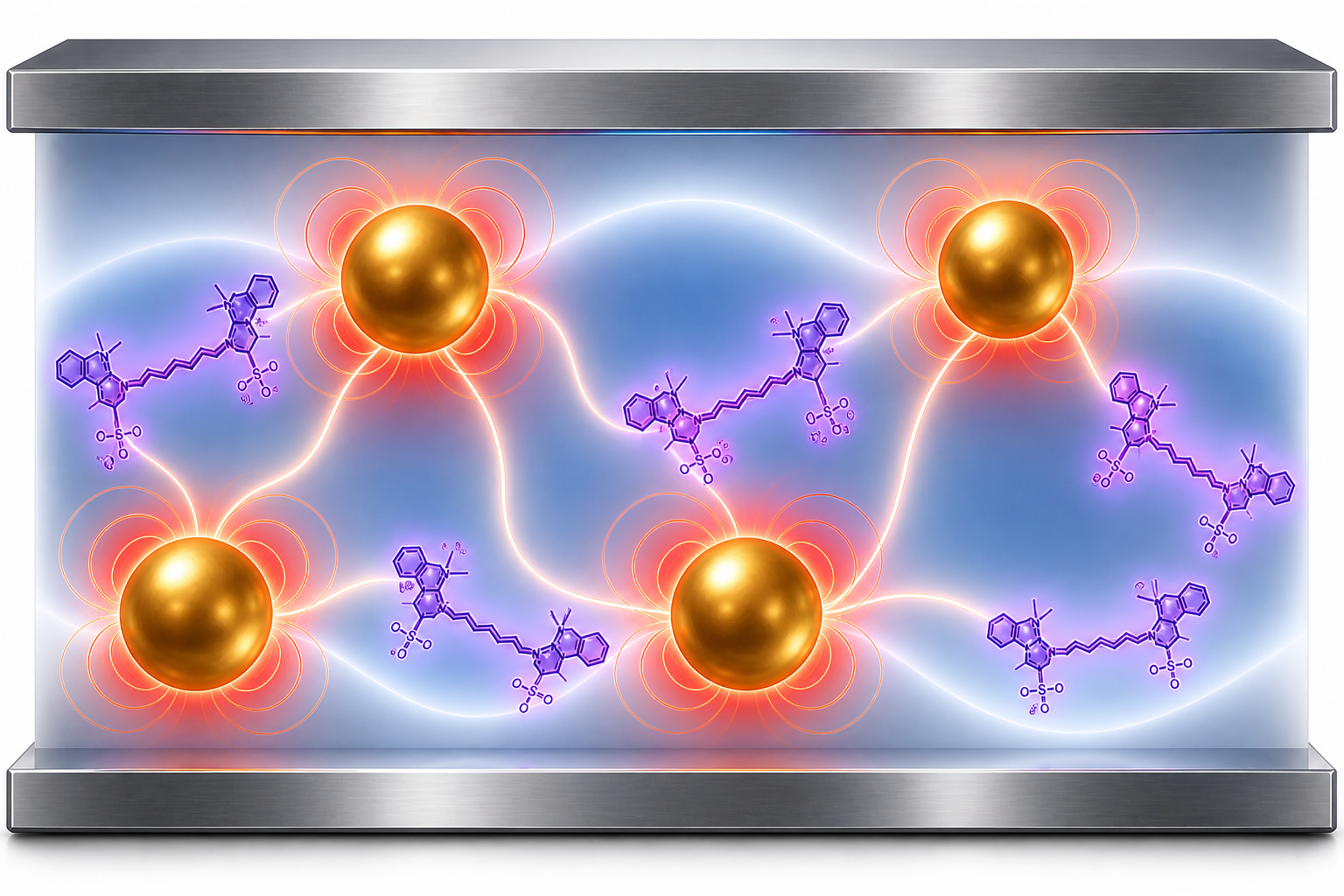}
	\end{center}
	
	\clearpage
	
	\begin{abstract}
		\noindent
		Strong coupling between molecules and electromagnetic fields has emerged as a powerful strategy to modify the physical and chemical properties of molecules, enabled by the formation of hybridized energy levels through strong light–matter interactions.
		Rather than purely photonic or plasmonic modes, hybrid cavity fields that integrate plasmonic and photonic contributions provide versatile platforms with distinct advantages for tailoring light–matter interactions.
		Here, we present hybrid modes formed through the coupling between localized surface plasmon resonances of Au nanoparticles and Fabry–Pérot microcavity modes.
		Furthermore, by using this hybrid platform, we demonstrate that the cavity field can give rise to cooperative coherent coupling between the spatially separated plasmonic nanoparticles and the molecules, by mediating long-range dipole-dipole interactions.
		As we show, this cavity-mediated coupling also enhances the plasmon–exciton mixing, as compared to their direct interaction.
		This configuration opens new avenues for tailoring light–matter interactions at the nanoscale, by supporting novel hybrid plexcitonic states that incorporate contributions from plasmons, excitons, and extended cavity fields.
	\end{abstract}
	
	\section*{Introduction}

	Strong coupling of organic molecules or semiconductor materials to the electromagnetic field of a resonant cavity can profoundly modify their photophysical and chemical properties.
	Such modifications have been the topic of vast research for several decades, and have driven the rapid emergence of the field of polaritonic chemistry in recent years.\cite {Ebbesen2016, Genet2021, Bhuyan2023}
	The strong coupling regime is characterized by the formation of hybrid quasiparticles, known as exciton-polaritons, which are coherent superpositions of photons and excitons, and whose wavefunctions can be delocalized over macroscopic molecular ensembles.
	While most studies on exciton polaritons employ planar Fabry-Perot microcavities to produce the confined optical mode, it is well known that metallic nanoparticles can also be utilized for strong light-matter coupling.
	In metallic nanoparticles, light couples to collective surface electron oscillations (plasmons) to produce localized surface-plasmon (LSP) modes.
	These modes are characterized by tightly confined electromagnetic fields that remain localized near the nanoparticle surface and decay into the surrounding medium.
	In a similar manner to cavity exciton-polaritons, the interaction between these localized plasmon modes and excitons can lead to the formation of the composite light–matter states (sometimes referred to as ``plexcitons'').\cite{Fofang2008,Fofang2011,Manjavacas2011,Zheng2017,Thomas2018}
	Due to the highly localized nature of plasmonic fields, the formation of plexcitonic states depends critically on the distance between the excitons and the metal nanoparticle surface.
	Therefore, plexcitonic systems normally necessitate the direct binding of molecules (or other excitonic materials) to the nanoparticle surface, which can be achieved using various strategies, such as electrostatic adsorption, covalent functionalization, layer-by-layer deposition, and bridging via short DNA linkers.\cite {Roller2016,Rodarte2017,Melnikau2019,Zhu2021,Ajaykumar2023}
	These approaches limit the choice of materials and often involve elaborate chemical modifications.
	Here, we take a radically different approach to couple the plasmonic modes with molecular excitons, even when the metal nanoparticles and molecules are spatially separated, by employing a hybrid plasmonic-photonic platform.

	Conventional strongly-coupled systems typically rely on purely photonic or plasmonic resonators.
	Although both platforms enable the formation of hybrid light–matter states, they are inherently constrained: photonic modes offer high quality factors but weak spatial confinement, whereas plasmonic modes provide strong confinement with nano-scale mode volumes but suffer from significantly higher losses.
	Recently, hybrid platforms that couple photonic and plasmonic modes have been proposed as a versatile alternative that offers mixed features, potentially mitigating the inherent drawbacks of each counterpart separately.
	Such structures have been used, for instance, to enable long-distance energy transfer, modify molecular reactivity, and enhance light–matter interactions.\cite {Akulov2018,BenAsher2025,Bisht2019,Hertzog2021} 
	
	In this work, we first demonstrated the coupling between the standing-wave modes of a Fabry-Perot cavity and localized surface plasmons in gold nanoparticles (AuNPs), randomly dispersed inside the cavity, leading to the formation of hybrid (plasmon-photon) polaritons.
	By analyzing the dispersion curves of these polaritons, we systematically investigated the extent and nature of hybridization between plasmonic and photonic modes, resulting in the emergence of delocalized polaritonic states from highly localized plasmon excitations.
	We then proceed to show how this hybrid plasmonic-photonic platform can be further coupled with molecular excitons to produce novel ``cavity-plexciton'' states that consist of contributions from excitonic states, localized plasmons, and delocalized photonic modes.
	Notably, under identical densities of AuNPs and molecules but in the absence of the cavity, the spatial separation between them hinders strong coupling between the plasmons and excitons. 
	As we show, in such a configuration, the cavity mediates long-range dipole-dipole interactions between the plasmonic nanoparticles and molecules, which leads to their coherent coupling despite being spatially separated by tens of nanometers.
	These results demonstrate a novel strategy to enable light–matter hybridization between otherwise uncoupled plasmons and excitons via a cavity field. 
	Furthermore, this hybrid plasmon–cavity platform and cavity–plexciton quasiparticles provide a versatile framework for tailoring light–matter interactions, with potential applications in plasmon-assisted reactions and nanoscale energy-transfer processes.

\begin{figure}[b!]
		\centering
		\includegraphics[width=8.25cm]{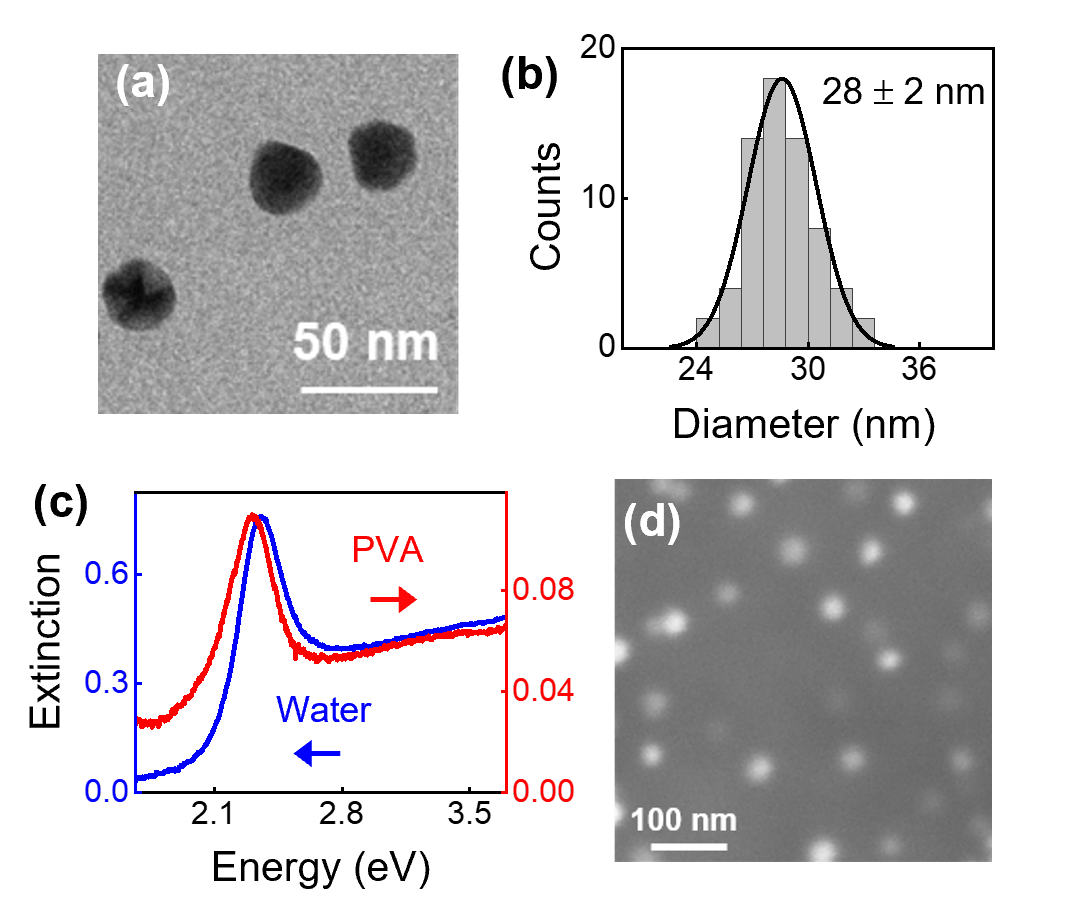}
		\caption{(a) Representative TEM image of the as-synthesized gold nanoparticles (AuNPs). (b) Size histogram of the AuNPs, calculated for an ensemble of $\sim$70 particles (see Supporting Information). (c) Extinction spectra of AuNPs in water (blue trace) and embedded in a PVA polymer matrix (red trace). (d) HRSEM image of AuNPs embedded in the PVA matrix deposited on a 35 nm Ag film.}
		\label{fig:AuNP}
	\end{figure}

	Our hybrid photonic--plasmonic cavity employs a metallic Fabry--Perot microcavity made of two Ag mirrors, separated by a polyvinyl alcohol (PVA) spacer which is in turn embedded with gold nanoparticles (AuNPs) (see Supporting Information for full details).
	The AuNPs were synthesized in water via a kinetically controlled seed-mediated growth protocol reported previously \cite {Bastus2011} (Supporting Information).
	Transmission electron microscopy (TEM) analysis reveals highly monodispersed AuNPs, with an average diameter of 28 ± 2 nm, as shown in \autoref{fig:AuNP}a,b. 
	Furthermore, the extinction spectrum of the AuNPs (blue curve in \autoref{fig:AuNP}c) exhibits a well-defined LSP resonance at 2.35 eV in water, and the narrow linewidth (233 meV FWHM) further confirms the high size-uniformity of the nanoparticles.
	To obtain the highest cooperative coupling to the cavity while still avoiding direct near-field interaction between the AuNPs, their concentration in the PVA matrix was optimized at $1.95\,\mathrm{\mu M}$ (number density $n \simeq 1.2 \times 10^{21}\,\text{m}^{-3}$), which was found to be the maximal concentration before aggregation occurs.
	As seen in \autoref{fig:AuNP}c (red curve), at this optimized concentration the spectrum of PVA/AuNPs film exhibits a slight redshift of the plasmon peak to 2.30 eV due to the higher refractive index of the host medium.
	Otherwise, it maintains its spectral shape and width, indicating that the AuNPs indeed remain isolated within the film.
	This is further confirmed by the high-resolution scanning electron microscopy (HRSEM) images of the PVA/AuNPs film (\autoref{fig:AuNP}d), deposited on top of a 35 nm-thick Ag layer under the same conditions used when preparing the cavities below, showing that clustering hardly occurs.
	Additionally, TEM images of the PVA/AuNPs matrix provided in Supporting Information corroborate these observations, revealing a homogeneous distribution of well-isolated AuNPs throughout the PVA layer.
	
\begin{figure} [t]
		\centering
		\includegraphics[width=8.25cm]{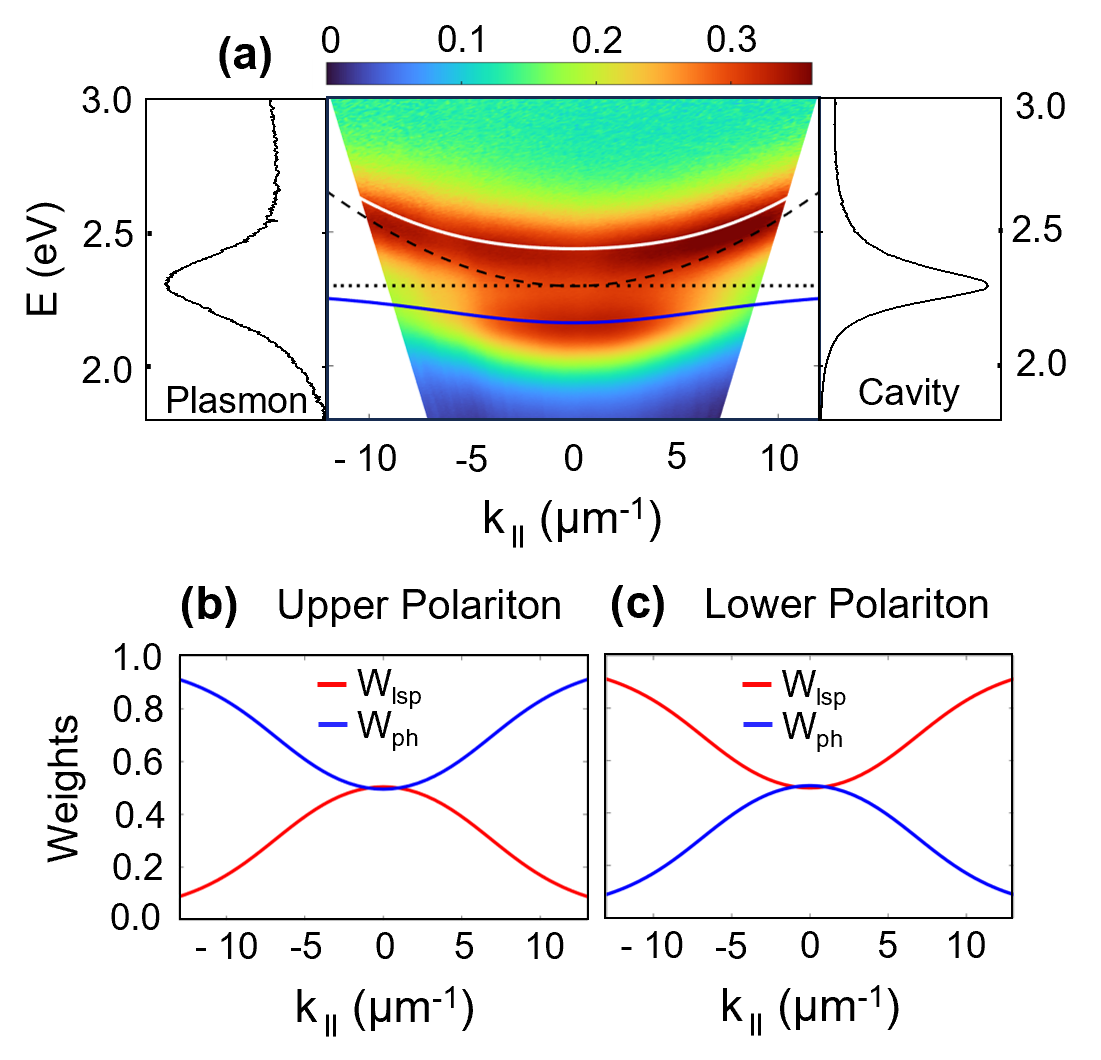}
		\caption{
			(a) Polariton dispersion of the hybrid cavity–plasmon system measured by angle-resolved reflection spectroscopy in TE polarization (false-color map). The solid blue and white lines correspond to the dispersion fitting of the lower and upper polariton branches, respectively, based on the coupled harmonic oscillator (CHO) model. The black dotted and dashed lines correspond to the localized plasmon and delocalized cavity modes, respectively. The left panel shows the measured extinction spectrum of the bare Au nanoparticles, while the right panel shows the transmission spectrum of the empty cavity.
			(b) Composition of the upper polariton branch of hybrid cavity–plasmon system, showing the plasmonic (red) and photonic (blue) weights as a function of in-plane momentum.
			(c) Same as (b), but calculated for the lower polariton branch.
		}
		\label{fig:Cavity-AuNP}
	\end{figure}

\begin{figure}
		\centering
		\includegraphics[width=0.5\linewidth]{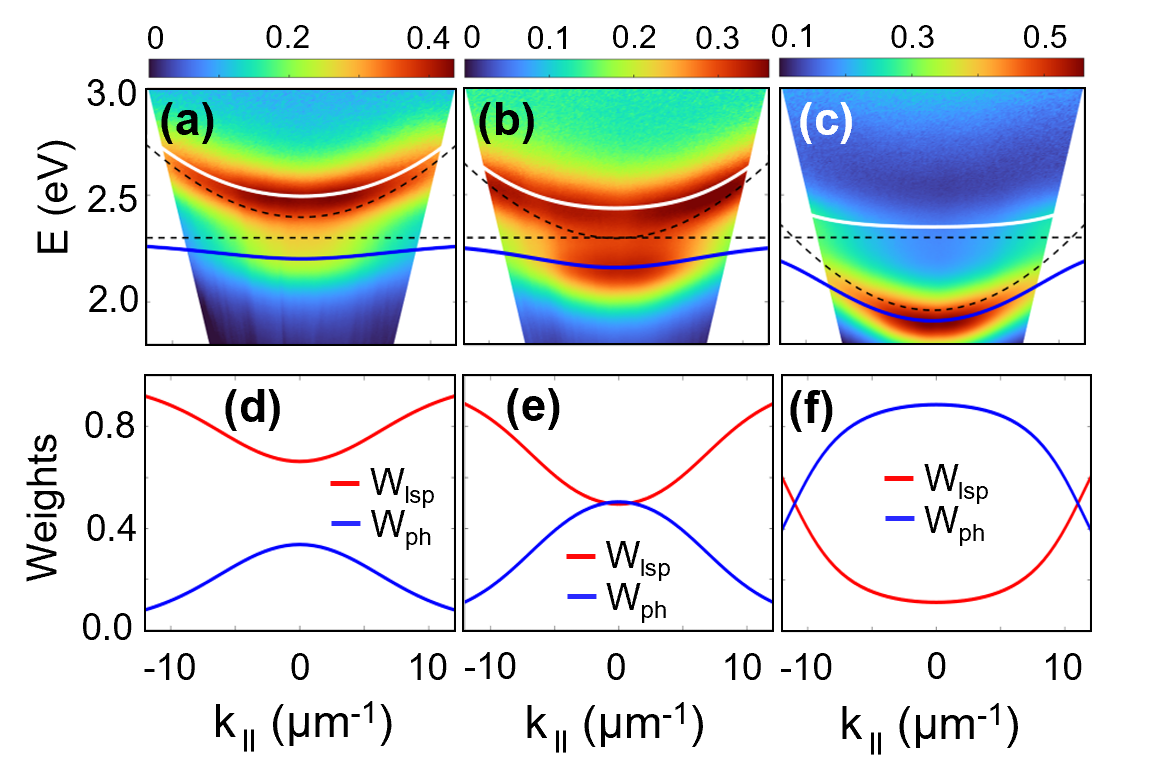}
		\caption{
			(a-c) Polariton dispersion diagrams for various hybrid systems with different cavity thicknesses, resulting in blue-detuned (a), tuned (b), and red-detuned (c) normal-incidence cavity modes (with respect to the plasmon resonance energy). The false-color images correspond to the measured dispersion while the solid lines correspond to fitting of the upper (white) and lower (blue) polariton branches to the CHO model.
			The horizontal and curved black dashed lines mark the localized plasmon energy and the dispersion of the delocalized cavity mode, respectively.
			(d-f) Polariton compositions along the lower branch, calculated for the three cavities in (a-c), showing the plasmonic (red) and photonic (blue) weights.
		}
		\label{fig:Detuning_Cavity}
	\end{figure}

	We proceeded to characterize the hybrid cavity, formed by placing the composite PVA/AuNPs film between two Ag mirrors (see Supporting Information).
	The mirror separation ($\simeq 150~\text{nm}$, set by the film thickness) was tuned such that the cavity mode under normal incidence was resonant with the localized plasmonic mode at 2.30 eV.
	As seen in \autoref{fig:Cavity-AuNP}a, the measured dispersion of the sample (acquired by angle-resolved reflectance spectroscopy in TE polarization, see Supporting Information) clearly displays two dispersive branches, residing above and below the LSP energy level (marked by the horizontal black dotted line).
	This avoided-crossing behavior is typical of strong coupling in Fabry-Perot microcavities, and in the present case it shows how the localized plasmons and the delocalized cavity mode hybridize to form two sets of eigenstates, corresponding to the upper and lower polaritons.
	In addition, we successfully fitted the measured dispersion curves with the coupled harmonic oscillator (CHO) model.
	In \autoref{fig:Cavity-AuNP}a, the black lines correspond to the bare plasmon resonance (dotted line) and the dispersive empty-cavity mode (dashed line), while the white and blue solid lines represent the upper and lower polaritonic branches resulting from the CHO model (see Supporting Information).
	From this fit we extract a coupling strength of $\hbar\Omega = 280~\text{meV}$ for the coupled system.
	This value indeed satisfies the strong-coupling criterion,
	$\hbar\Omega > (\Gamma\sub{lsp} + \Gamma\sub{c})/2$,
	where $\Gamma\sub{lsp} = 260~\text{meV}$ and 
	$\Gamma\sub{c} = 170~\text{meV}$ denote the 
	full widths at half maximum (FWHM) of the LSP in PVA and the cavity modes, respectively (see left and right insets of \autoref{fig:Cavity-AuNP}a).
	Finally, we used the CHO model to calculate the Hopfield coefficients and to obtain the composition of each polaritonic state, namely the delocalized photonic ($W\sub{ph}$) and localized plasmonic ($W\sub{lsp}$) weights, as a function of in-plane momentum $k_\parallel$ (see Supporting Information).
	The results are presented in \autoref{fig:Cavity-AuNP}b,c, with the photonic and plasmonic contributions represented by blue and red curves, respectively.
	At $k_\parallel = 0$ (corresponding to normal incidence, and for this sample zero detuning), both polariton branches exhibit nearly equal admixtures of photonic and plasmonic components, consistent with maximum hybridization. 
	At large in-plane momentum values, the upper polariton branch (\autoref{fig:Cavity-AuNP}b) becomes increasingly photonic in character, while the lower polariton branch  (\autoref{fig:Cavity-AuNP}c) acquires a predominantly plasmonic nature.
	To study the plasmon-cavity hybridization behavior in more detail, we fabricated hybrid cavities with two different thicknesses, such that cavity mode at normal incidence was either blue- or red-detuned from the localized plasmon resonance, enabling controlled tuning of the photonic weight in the lower polariton, results are summarized in \autoref{fig:Detuning_Cavity}.
	We repeated the CHO analysis for those two detuned samples with the same coupling strength, and successfully fitted the dispersion measurements (\autoref{fig:Detuning_Cavity}a-c).
	We stress that the same coupling strength was used for all three cavities, with the cavity thickness being the only fitting parameter that was varied.
	Furthermore, from the CHO calculations we obtained the composition of the polaritonic states in a similar manner to the tuned cavity, with the results (for the lower branch) shown in \autoref{fig:Detuning_Cavity}d-f for the blue-detuned cavity, the tuned cavity, and the red-detuned cavity, respectively.

	The formation of hybrid plasmon-photonic modes raises interesting possibilities for light-matter interaction, in particular for plexcitonic systems, where the nano-scale confinement of the localized plasmon modes restricts their interaction with quantum emitters to the immediate surrounding of the metal nanoparticles' surface.
	The dispersive nature of the hybrid plasmonic-photonic modes, as observed in \autoref{fig:Cavity-AuNP} and \autoref{fig:Detuning_Cavity}, indicates that the strong coupling between the nanoparticles and the cavity field transforms the localized plasmonic excitation into a delocalized (spatially extended) state.
	This, in turn, can extend the effective interaction volume, enabling cavity-mediated coupling between the plasmons and distant emitters.

\begin{figure}[t]
		\centering
		\includegraphics[width=8.25cm]{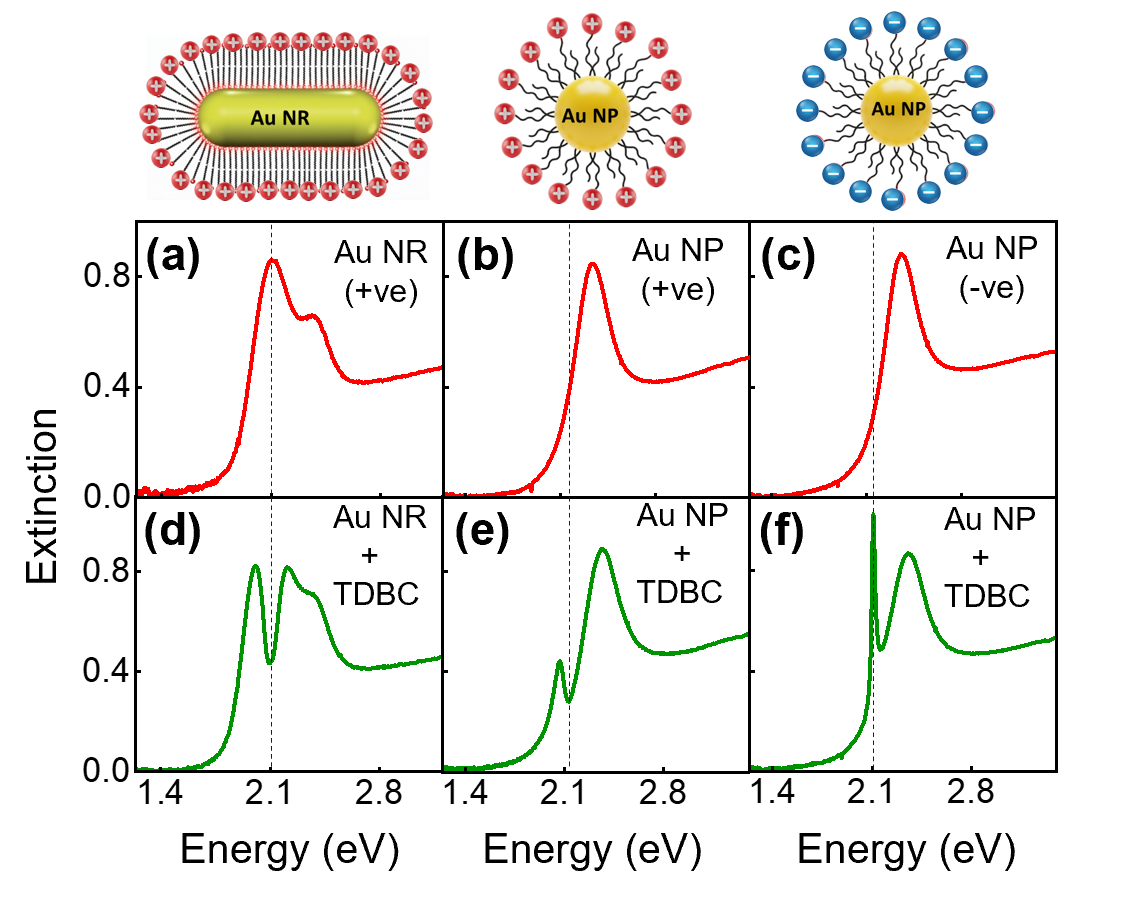}
		\caption{
			Comparison of the coupling between various plasmonic nanoparticles and TDBC J-aggregates molecules, showing the measured extinction spectra for the bare nanoparticle solutions (a-c), and after mixing with the TDBC solution (d-f).
			Positively charged AuNRs (a) electrostatically bind TDBC, leading to resonant interaction between the longitudinal plasmon mode and the excitons and spectral splitting for the hybridized system (d).
			Positively charged AuNPs (b) also bind TDBC, but have detuned plasmon mode, leading to spectral splitting with uneven polariton peaks (e).
			Negatively charged AuNPs (c) repel the TDBC molecules, hindering the near-field interaction between them. Consequently, the extinction spectrum of the mixed solution (f) is a simple addition of the AuNPs and TDBC spectra.
			In all graphs the vertical dotted lines mark the absorption maximum of the TDBC J-aggregates.
		}
		\label{fig:Plexciton}
	\end{figure}

	Before proceeding to demonstrating this effect, we first examine plexciton formation in three simple nanoparticle geometries, for comparison.
	As summarized in \autoref{fig:Plexciton}, we consider gold nanorod (AuNR) with positive surface charge, and spherical gold nanoparticles (AuNP) with either positive or negative surface charge (see Supporting Information).\cite{Ye2012,Bastus2011}
	Upon mixing the positively charged AuNRs (with longitudinal and transverse plasmon modes at 2.11 and 2.36\,eV, respectively, see \autoref{fig:Plexciton}a) with negatively charged TDBC J-aggregates in aqueous solution, the molecules are electrostatically adsorbed on the AuNR surface (for further details see Supporting Information).
	This system satisfies the two essential criteria for efficient plexciton formation: spectral resonance between the plasmon mode and the exciton energy (at 2.11\,eV) and close spatial proximity, which allows the near-field interaction between them.
	As seen in \autoref{fig:Plexciton}d, this on-resonant interaction results in the splitting of the longitudinal plasmon extinction peak, marking the formation of an upper ($P^{+}$) and lower ($P^{-}$) plexcitonic states at 2.21~eV and 2.00~eV, respectively, with equal plasmonic and excitonic weights for both polaritons.
	In contrast, the large detuning ($\sim 250$~meV) between the transverse plasmon resonance with the exciton energy prevents efficient mixing between them, and therefore the transverse mode remains almost unaffected at 2.36\,eV.
	Assuming a similar coupling strength as for the longitudinal mode (i.e., a Rabi frequency of 210\,meV), this mode has an excitonic weight of 0.10 at most, as estimated from the CHO model.
	Next, to investigate the role of resonance energy in plexciton hybridization, we replaced the AuNRs with spherical, positively charged AuNPs, for which the plasmon resonance at 2.33~eV (\autoref{fig:Plexciton}b) is slightly detuned from the exciton energy.
	When mixing the AuNPs with the J-aggregate solution, the coupling between the plasmon and the excitons again leads to spectral splitting, as seen in \autoref{fig:Plexciton}e.
	The peaks at 2.36 and 2.06~eV again correspond to an upper ($P^{+}$) and lower ($P^{-}$) plexciton states, however, due to the 230~meV detuning between the plasmon and the exciton, the lower polariton becomes predominantly excitonic (with an exciton weight $W^{(LP)}\sub{x} = 0.86$ and plasmonic weight $W^{(LP)}\sub{lsp} = 0.14$), whereas the upper polariton is primarily plasmonic ($W^{(UP)}\sub{lsp} = 0.86$, $W^{(UP)}\sub{x} = 0.14$), as obtained from the CHO model, which also gives a coupling strength of 210\,meV (see Supporting Information).
	This occurs because, although the AuNPs and the molecules are in close proximity, the detuning between the plasmon and exciton energies reduces the efficient mixing of these two components in the resulting plexcitonic states. 
	Finally, we repeated these measurements with negatively charged AuNPs, again mixing them with the TDBC J-aggregates solution. 
	In contrast to the positively charged AuNRs and AuNPs, in this case electrostatic repulsion prevents the molecules from entering the nanometer-volume around the nanoparticles, resulting in complete suppression of their interaction with the localized plasmonic field.
	Therefore, this system fails to meet the essential conditions for plexciton formation, lacking both spectral resonance and close spatial proximity between the constituents.
	Therefore, no hybridization can occur, as can be seen by comparing the spectra of the AuNP solution (\autoref{fig:Plexciton}c) and the mixed solution (\autoref{fig:Plexciton}f), exhibiting a simple additive sum of the uncoupled plasmonic and excitonic contributions.
	
	Overall, the results presented in \autoref{fig:Plexciton} illustrate the well-known importance of both near-field coupling and spectral tuning between the interacting species, for efficient plasmon–exciton hybridization.
	On the other hand, as we demonstrated above, when placed inside a planar microcavity, the localized plasmons can be hybridized with the cavity field to produce extended, spectrally-shifted modes with significant plasmonic contributions.
	The partial delocalization of these modes suggests that they may be coupled with molecular excitons, even if the molecules are not in close proximity to the AuNPs.
	In other words, instead of direct near-field interaction between molecules and the AuNPs, the cavity mode, by virtue of its coherent coupling with both the excitons and the LSPs, mediates long-range dipole-dipole interactions between them.
	Such a mechanism is reminiscent of the cavity-mediated, long-range intermolecular energy transfer that was observed and studied over the past few years \cite{Coles2014,Zhong2017,Sandik2025,Kertzscher2026}.
	In a similar manner to those strongly-coupled systems, where polaritonic states with both donor and acceptor are formed, the hybrid plasmon-photonic platform can be used to form polaritonic modes with significant plasmonic and excitonic contributions, even when the conditions discussed above are not met.

\begin{figure}[t]
		\centering
		\includegraphics[width=8.25cm]{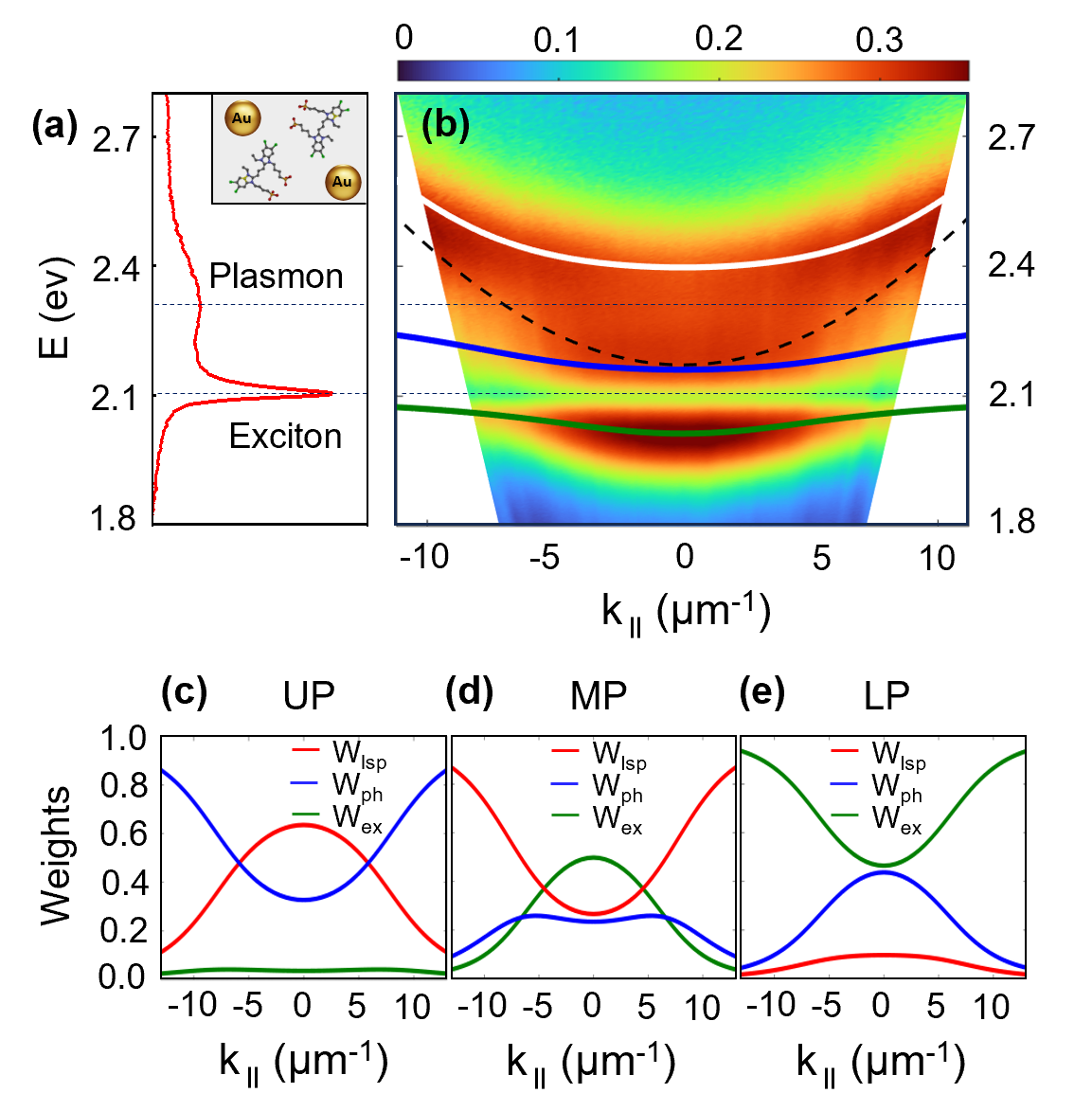}
		\caption{(a) Extinction spectrum of a PVA host film embedded with AuNPs and TDBC J-aggregates. (b) Polariton dispersion of the hybrid cavity–plexciton system, measured under TE polarization using angle-resolved reflection microscopy (false-color map). The solid white, blue and green lines correspond to the calculated dispersion curves of the upper, middle and lower polariton branches, respectively, based on the coupled harmonic oscillator (CHO) model.
			The black dashed line corresponds to the dispersion of the empty cavity mode, and the horizontal black dotted lines mark the energies of plasmons and excitons. (c,d,e) Calculated plasmonic (red line), photonic (blue) and excitonic (green) weights for the upper (c), middle (d), and lower (e) polariton states in the hybrid cavity–plexciton system.}
		\label{fig:Cavity-polariton}
	\end{figure}

	To demonstrate this cavity-mediated long-range interaction, we fabricated a Fabry–Pérot microcavity filled with AuNPs and TDBC J-aggregates, both uniformly dispersed within the cavity volume.
	Prior to cavity fabrication, a PVA film embedding both constituents was prepared and characterized outside of the cavity.
	The extinction spectrum of the resulting composite polymer film (\autoref{fig:Cavity-polariton}a) exhibits two distinct absorption features at 2.30 and 2.11~eV, corresponding to the LSP resonance of the AuNPs (as in \autoref{fig:AuNP}c) and the excitonic transition of the TDBC J-aggregates, respectively.
	The absence of spectral distortion, linewidth modification, or mode splitting confirms that direct plasmon–exciton interactions are negligible under these conditions, such that the absorption of the film is given as a simple addition of the absorption spectra of the two species.
	Next, we fabricated a Fabry–Pérot microcavity consisting of two Ag mirrors and the composite film between them, maintaining the same concentrations as in \autoref{fig:Cavity-polariton}a for the AuNPs and TDBC J-aggregates within the PVA matrix (see Supporting Information).
	We adjusted the film thickness to $L\sub{cav}\simeq 163~\text{nm}$, such that the cavity resonance at normal incidence lies between the plasmon resonance and exciton energy levels, enabling efficient coupling with both constituents.

	We further characterized the resulting hybrid system using angle-resolved reflection spectroscopy (see Supporting Information). 
	Under this resonance condition, the interaction between localized plasmons, excitons, and the cavity field generates three hybrid energy levels at 2.01, 2.16, and 2.39 eV (at normal incidence), corresponding to the lower, middle, and upper polariton states, respectively.
	The corresponding dispersion map is shown in \autoref{fig:Cavity-polariton}b (false-color plot).
	In a similar manner to cavities filled with two molecular species, the extracted dispersion curves were successfully fitted using a three-oscillator model\cite{Coles2014,Zhong2016,Dutta2025}, having the form
	\begin{equation}
		H = 
		\begin{pmatrix}
			E\sub{c}(k_\parallel) & g\sub{cp} & g\sub{cx} \\
			g\sub{cp} & E\sub{lsp} & 0 \\
			g\sub{cx} & 0 & E\sub{x}
		\end{pmatrix}
		\label{eq:CHO}
	\end{equation}
	where $E\sub{c}(k_\parallel)$, $E\sub{lsp}$, and $E\sub{x}$ denote the (dispersive) cavity mode, localized plasmon, and exciton energies, respectively.
	Here, we account for the cavity–plasmon interaction (quantified by $g\sub{cp}$) and the cavity–exciton interaction (with the coupling constant $g\sub{cx}$), while neglecting direct plasmon–exciton coupling.
	We take $g\sub{cp} = 140\,\mathrm{meV}$ (based on the results in \autoref{fig:Cavity-AuNP} above) and obtain $g\sub{cx} = 100\,\mathrm{meV}$ from an independent measurement of an equivalent TDBC-cavity system (see Supporting Information).
	As seen in \autoref{fig:Cavity-polariton}b, the eigenenergies of this Hamiltonian (solid lines) perfectly match the experimental measurement, with the effective cavity length being the only fitting parameter, indicating that this system indeed leads to a tripartite hybridization between the plasmons, excitons and cavity mode.

	Finally, using the fitted $3\times3$ CHO model, we calculated the Hopfield coefficients of the hybrid system to obtain the composition of each polaritonic state and its dependence on the in-plane momentum (see Supporting Information).
	The results are presented in \autoref{fig:Cavity-polariton}c-e for the upper (c), middle (d), and lower (e) polariton branches, and with the photonic ($W\sub{ph}$), excitonic ($W\sub{x}$) and plasmonic ($W\sub{lsp}$) weights represented by blue, green and red curves, respectively. 
	As can be seen, the upper polariton branch is dominated by plasmonic and photonic contributions, with $W^\text{(UP)}\sub{lsp} = 0.64$ and $W^\text{(UP)}\sub{ph} = 0.33$ at normal incidence ($k_\parallel = 0$), while the excitonic contribution remains negligible ($W^\text{(UP)}\sub{x} = 0.03$).
	In other words, the upper polariton branch has predominantly electromagnetic character.
	In sharp contrast, the middle polariton branch exhibits significant contributions from all three counterparts across most of the measured angular range.
	Specifically, at $k_\parallel = 4.52~\mu m^{-1}$, where the cavity resonance resides nearly midway between the plasmon and the exciton energy (slightly to plasmon), the middle polariton branches has nearly equal photonic ($W^\text{(MP)}\sub{ph} = 0.26$), plasmonic ($W^\text{(MP)}\sub{lsp} = 0.37$) and excitonic ($W^\text{(MP)}\sub{x} = 0.37$) weights.
	The near-equal plasmonic and excitonic contributions signify the cavity-mediated coherent interaction between the TDBC molecules and the AuNPs which leads to hybridization of their excitations, since, outside the cavity the plasmonic and excitonic degrees of freedom are completely noninteracting.
	In line with the notion of plexcitons, which form under the direct coupling between localized plasmons and excitons, we refer to such hybrid states, where the hybridization between plasmons and excitons is mediated by the cavity field, as ``cavity-plexcitons''.
	Interestingly, we also find that the cavity-mediated interaction can also effectively bridge over the energetic gap between the plasmons and the excitons.
	As discussed in \autoref{fig:Plexciton}b and e, in the coupled AuNP-TDBC system with a coupling strength of 210~meV, the plasmon-exciton detuning results in coupled states which display moderate light-matter hybridization, and are close to being either purely plasmonic or purely excitonic ($W\sub{x} = 0.86$  and $W\sub{ph} = 0.14$ for the lower polariton and vice versa for the upper one).
	However, as pointed above, the cavity-mediated interaction leads to a mid-polariton state with which is 37\% plasmonic and 37\% excitonic.
	In other words, even though the coupling strength and the plasmon-exciton detuning are similar, this system exhibits efficient and more balanced mixing between them.
	Finally, inspecting the various weight distributions in the lower polariton branch, we observed that it exhibits a pronounced exciton--cavity hybridization, with nearly equal excitonic and photonic weights at $k_\parallel = 0$ ($W^\text{(LP)}\sub{x} = 0.46$, $W^\text{(LP)}\sub{ph} = 0.44$) and a rather small plasmonic contribution, with $W^\text{(LP)}\sub{lsp} = 0.1$.
	As such, these modes resemble the usual cavity-polaritons that form in organic cavities.
	
	\begin{figure}
		\centering
		\includegraphics[width=8.25cm]{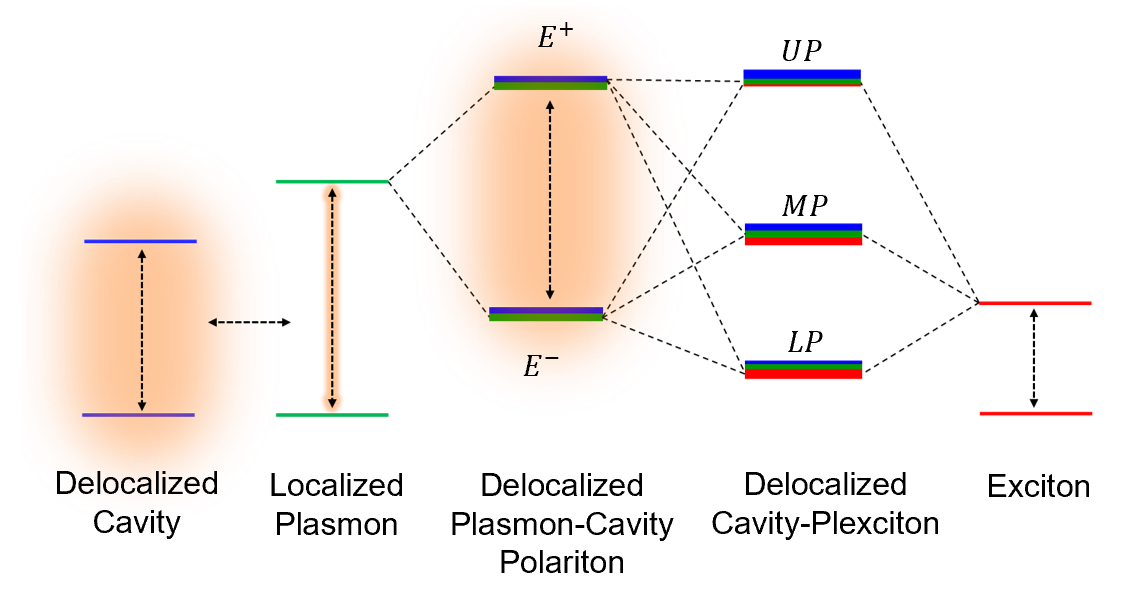}
		\caption{Schematic representation of cavity--plexciton formation. Note: localized plasmons (AuNPs) and excitons (TDBC J-aggregates) are uncoupled in free space.}
		\label{fig:CavPlexcitonScheme}
	\end{figure}
	In the analysis above we followed the usual, straightforward procedure and directly diagonalized the $3\times3$ CHO Hamiltonian (\autoref{eq:CHO}) to calculate the polariton dispersion diagram and compositions.
	However, to gain further insight into the mechanistic pathways underlying the formation of these cavity-plexcitons, it is instructive to represent the interactions in this system by a two-step process.
	As illustrated schematically in \autoref{fig:CavPlexcitonScheme}, we first account for the interaction between the AuNPs and the cavity mode, which results in the formation of two hybrid plasmonic-photonic modes, as demonstrated above.
	By transforming the Hamiltonian using such partial diagonalization (see full details in Supporting Information), we obtain the intermediate Hamiltonian
	\begin{equation}
		H =\begin{pmatrix}
			E^{+}   & 0     & g^{(+)} \\
			0       & E^{-} & g^{(-)} \\
			g^{(+)}   & g^{(-)} & E\sub{x}
		\end{pmatrix}
		\label{eq:CHO_Intermediate}
	\end{equation}
	with $E^{+} = 2.39~\text{eV}$ and $E^{-} = 2.08~\text{eV}$ being the energies of the upper and lower hybrid modes, respectively.
	As shown in \autoref{fig:Cavity-AuNP} above, for cases where the cavity resonance is red-detuned with respect to the LSP mode, the upper polariton has a larger plasmonic weight, while the lower polariton is more photonic.
	Specifically, when the cavity energy resides exactly between $E\sub{lsp}$ and $E\sub{x}$, we obtain the photonic/plasmonic weights as $W^{(+)}\sub{ph} = 0.29$ and $W^{(+)}\sub{lsp} = 0.71$ for the upper mode and $W^{(-)}\sub{ph} = 0.71$ and $W^{(-)}\sub{lsp} = 0.29$ for the lower one (see Supporting Information).
	In this picture, the hybrid plasmonic–photonic modes subsequently couple to the molecules to produce the upper ($E\sub{UP}$), middle ($E\sub{MP}$), and lower ($E\sub{LP}$) cavity-plexciton states, as discussed above.
	However, due to the different compositions of the plasmonic-photonic modes, each one of them exhibits different interaction strength with the excitons, which is represented by the two \textbf{effective coupling constants} $g^{(+)} = 54~\text{meV}$ and $g^{(-)} =84~\text{meV}$ (see Supporting Information).
	Notably, due to the smaller effective coupling constant and the large detuning between $E^{+}$ and $E\sub{x}$ ($\sim 280$~meV), the energy of the upper cavity–plexciton is practically unchanged from the plasmon-cavity polariton by the presence of the molecules, remaining at 2.39~eV.
	Likewise, its composition is hardly modified, and it acquires a negligible excitonic weight of $W^{\text{UP}}\sub{x} = 3.7\%$.
	In sharp contrast, the energy of the lower hybrid state ($E^{-}$ in \autoref{fig:CavPlexcitonScheme}) is pushed closer into resonance with the exciton, in addition to its larger coupling constant $g^{(-)}$.
	As a result, this hybrid mode exhibits efficient mixing with the excitons, producing a middle and a lower cavity-plexciton states which are 50\% and 46\% excitonic, respectively, at normal incidence (see \autoref{fig:Cavity-polariton} and Supporting Information).
	It is interesting to note that the hybrid system we have presented shares some similarities with surface-lattice resonances, where the localized plasmons in nanoparticle arrays couple to form hybrid plasmonic-photonic modes, which were also used for strong light-matter coupling.\cite{Rodriguez2013,Vakevainen2014,Wang2018,Patel2025}
	However, in our case, the molecular excitons are hybridized with plasmonic nanoparticles that are randomly distributed in space and therefore do not exhibit any direct, collective interactions.
	Instead, the collectivity in our system is only enabled through the coupling with the cavity mode, which in turn mediates the long-range interaction between the nanoparticle and molecular ensembles to create the hybrid cavity plexciton.
	
	In conclusion, we demonstrated the realization of a hybrid photon-plasmon electromagnetic field formed through the interaction between a Fabry–Pérot microcavity mode and the localized plasmons in Au nanoparticles which are randomly dispersed within the cavity volume. 
	This hybridization gives rise to novel polariton modes that mitigate the intrinsic limitations of each constituent by combining the highly localized plasmonic modes with the delocalized and comparatively low-loss nature of the cavity mode.
	We showed that plasmon–cavity polaritons originating from inherently localized plasmon resonances acquire an extended character, as indicated by their dispersive nature, while still retaining a pronounced plasmonic contribution. 
	This, in turn, extends the effective interaction volume, enabling cavity-mediated long-range coupling between the spatially separated plasmonic modes and molecular excitons, that otherwise remain uncoupled under free space conditions.
	This cavity-mediated coherent coupling leads to the formation of ``cavity-plexciton'' states, containing contributions from delocalized cavity photons, localized plasmons, and molecular excitons.
	Interestingly, under energetic detuning, we found that the cavity can effectively bridge the energy gap between the plasmons and excitons, further enhancing their mixing, as compared to their direct interaction in free space.
	Finally, we have introduced a two-step coupling model that can be used to qualitatively understand the hybridization in such composite, tripartite systems. 
	Beyond providing a new route to hybrid light–matter states, this platform establishes a versatile framework for engineering collective interactions, controlling energy redistribution at the nanoscale, and exploring hybrid properties in strongly coupled nanoplasmonic systems, polaritonic chemistry, and quantum plasmonics.
	
	\bibliography{Plasmon-Cavity-Exciton}
	
\end{document}